\documentclass[aip,reprint]{revtex4-1}
\usepackage{graphicx}
\usepackage{dcolumn}
\usepackage{bm}
\draft 

\begin{document}


\title{Mode surgery of LN micro-resonator by femtosecond laser irradiation} 



\author{Licheng Ge}
\author{Haowei Jiang}
\author{Yi'an Liu}
\author{Bing Zhu}
\author{Chenghao Lu}
\author{Yuping Chen}
 \email{ypchen@sjtu.edu.cn}
\author{Xianfeng Chen}
 \email{xfchen@sjtu.edu.cn}
\affiliation{State Key Laboratory of Advanced Optical Communication Systems and Networks, School of Physics and Astronomy, Shanghai Jiao Tong University, 800 Dongchuan Road, Shanghai 200240, China}


\date{\today}

\begin{abstract}
Lithium niobate(LN) microresonator is a key component in photonic integrated circuits. An intriguing phenomenon was observed that the quality factor for specific modes of the resonator can be further improved by femtosecond laser irradiation. The high repetition laser pulses scatters into the cavity by forming a defect on the surface and is confined in the edge. A localized coupled laser refining process will then repair the lattice disorders and tiny burrs on the periphery. The intrinsic quality factor of a measured mode can be promoted from 2.17$\times$10$^5$ to 1.84$\times$10$^6$ indicating one order of magnitude improvement.  This highly controllable method has its own unique advantages in improving the quality of an integrated resonator which opens new prospects for single component engineering of LN photonic integrated circuits.
\end{abstract}


\maketitle 

Lithium niobate (LN) crystal offers attractive properties such as electro-optic, nonlinear optical and acousto-optic effects in a wide wavelength range so that it is awarded as ``photonic silicon''. The commercialization of lithium niobate on insulator (LNOI) is revolutionizing the lithium niobate industry\cite{poberaj2012lithium}, calling significant interest in the platform for photonic intgrated circuits on a single chip\cite{boes2018status}. LNOI has excellent properties as bulk LN and at the same time offers stronger optical confinement and a high optical element integration density which has already made great success in silicon on insulator platform. As chip components, optical waveguide\cite{cai2015low,cai2015waveguides}, wavelength converter\cite{chang2016thin}, electro-optic modulator\cite{preble2007changing,wang2017chip} and resonant structures\cite{wang2012high,zhang2017monolithic} are all intensively studied in the past few years. Here we focus on whispering gallery mode (WGM) micro-disk resonators. It confines light at the edge of the resonator via total internal reflection leading to remarkably high intracavity optical intensities. It has attracted much attention also due to the combination of high quality factor (Q factor) and tunable coupling features. Utilization of electro-optic effect on micro-disk resonators produces fast and effective tuning of resonant wavelength, and thus the tunability on light intensity and filtered wavelength. The nonlinear frequency conversion can be greatly enhanced by using the micro-disk resonator. All of these has made it a suitable platform for fundamental physics studies such as nonlinear frequency conversion\cite{hao2017sum,lin2015second,breunig2016three}, electro-optic effects\cite{wang2015high,guarino2007electro,wang2018nanophotonic}, photorefractive effects\cite{jiang2017fast}, optomechanics\cite{jiang2016chip} and quantum physics.

\begin{figure}[b]
    \centering 
    \includegraphics[width=\columnwidth]{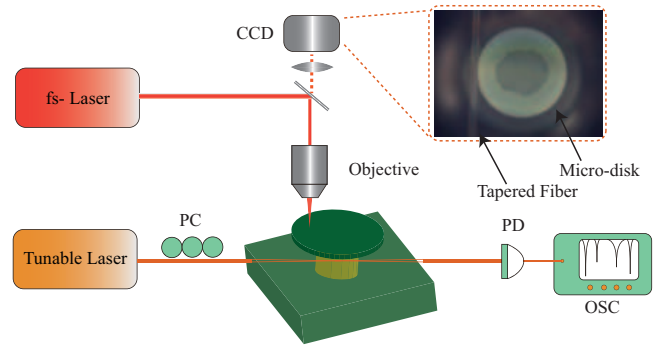}
    \caption{Experimental setup of the whole system. The orange line stands for the tunable probe laser while the red line stands for the interacting femtosecond laser. The inset shows the image taken from CCD. PC: polarization controller. PD: photodetector. OSC: Oscilloscope.}
    \label{fig1}
\end{figure}

The fabrication method of LN resonator has been intensively studied and different scales of micro-disks have been successfully produced. Mechanical polishing is a common way to produce millimeter scale LN disks and the Q factors can reach hundreds of millions\cite{ilchenko2004nonlinear}. It finds applications in many areas such as natural phase matching\cite{furst2010naturally}, electric field sensor and so on. For the demand of integration micro scale disks are developed using semiconductor fabrication methods though LN is a relatively hard crystal to be etched\cite{lacour2005nanostructuring}. The first LN micro-disk was fabricated using selective ion implantation combined with chemical etching and thermal treatment\cite{wang2012high}. After the commercialization of LN thin film, fabrication utilizing electron beam lithography and photolithography has taken the dominant position\cite{wang2014integrated,wang2015high} enabling mass production and being cost effective. A more flexible method combining femtosecond laser writing and focused ion beam (FIB) milling has been proposed which is more powerful and convenient to engineer single micro-disk\cite{lin2015fabrication}. In the most recent studies a chemo-mechanical polishing process was applied and the quality factors can be further improved\cite{wu2018lithium}. The most important factors to achieve high quality are clean surface and smooth sidewall\cite{guha2017surface}. Thermal treatment procedure is often added to remove burrs on the periphery. For silica microresonators high power CO$_2$ laser reflow is the most common way to do thermal treatment and can form a smooth toroid shape\cite{armani2003ultra,hossein2007free}. Another way is to put the sample in an annealing oven for several hours and the quality can have a significant improvement. However the entire integrated chip will experience the heat treatment process in both of the methods which may introduce some undesirable changes to other components on the chip. Here we demonstrate a new method to engineer the quality of a fabricated micro-disk in a localized and highly controllable way. By irradiating LN micro-disk with high repetition femtosecond laser pulses, we found that the quality factor for specific modes can have one order of magnitude increase in case a small defect is formed on the focusing spot. It is like a precision surgery that light is scattered into the disk via the defect and is confined in the periphery thus it will have strong interaction with lattices in the edge of LN disk. A coupled laser refining (CLR) process that mainly contains nonlinear absorption and resolidification process may contribute most to the quality improvement.

\begin{figure}[b]
    \centering 
    \includegraphics[width=\columnwidth]{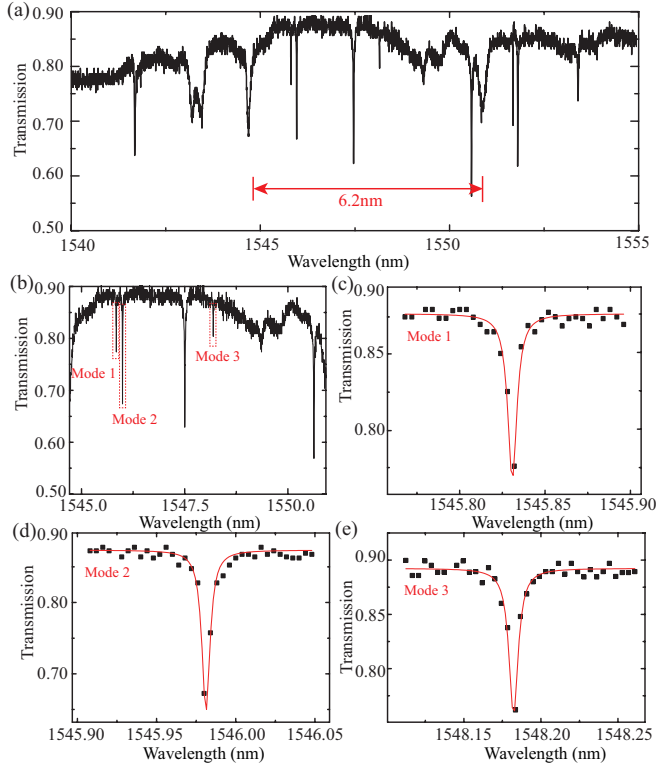}
    \caption{(a)Transmission spectrum of the LN micro-disk resonator with a diameter of 55$\mu$m. (b) Magnified one free spectral range in the whole spectrum. Three high Q modes are marked with labels. (c) (d) and (e) show the Lorentzian fitting of the labeled modes.}
    \label{fig2}
\end{figure}

Experimental setup is shown in Fig.$\ref{fig1}$. The light from C-band tunable laser serves as probe light and passes through a polarization controller in order to control the polarization state. Then it is coupled to the tapered fiber for exciting the resonant modes of the micro-disk. The spectrum of the output light is measured by oscilloscope (OSC).  Another laser is femtosecond laser which plays the role of interacting wave which goes through a micromaching system and is focused on the surface via a 100$\times$ lens. The sample is put on a precision XYZ piezo-stage which is used to control the distance between micro-disk and tapered fiber. Both of them are placed on the moving stage of the micromaching system. The inset shows the optical image of a coupled micro-disk system.

 We fabricated the micro-disk using a FIB assisted femtosecond laser direct writing method\cite{lin2015fabrication}. First we used femtosecond laser to inscribe the prototype on a 600nm LNOI film (from NANOLN) and then FIB was applied to polish the periphery. The silica layer underneath the disk is partially removed by hydrofluoric etching, only leaving a silica pedestal in the disk center behind. Fig.$\ref{fig2}$(a) shows the original transmission spectrum. The free spectral range (FSR) was measured to be 6.2nm and in order to analyze the impact of femtosecond laser interaction we focus on all the high Q modes marked as 1, 2 and 3 in a FSR as depicted in Fig.$\ref{fig2}$(b). The detailed spectrum with Lorentzian fit (Fig.$\ref{fig2}$ (c) (d) and (e)) show high intrinsic Qs of 2.65$\times$10$^5$, 3.00$\times$10$^5$ and 2.17$\times$10$^5$.

\begin{figure}[t]
    \centering 
    \includegraphics[width=0.95\columnwidth]{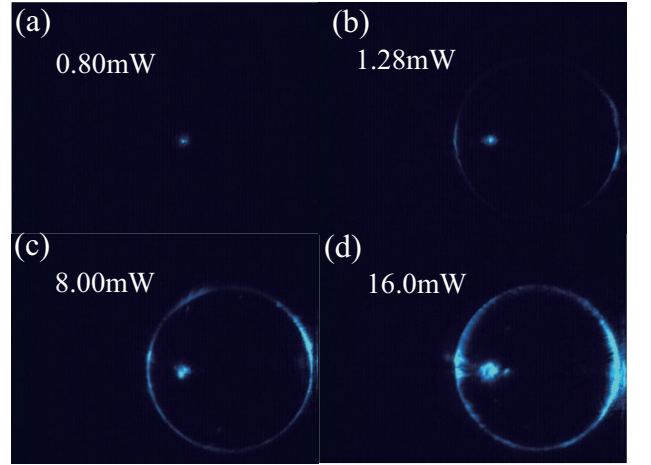}
    \caption{Image of irradiation of femtosecond laser on LN micro-disk resonator when laser power was set at (a)0.80mW, (b)1.28mW, (c)8.00mW and (d)16.0mW}
    \label{fig3}
\end{figure}

Then we turn on the femtosecond laser and focus it on the surface of the micro-disk. The laser is centered at 1030nm with a pulse width of 600fs and the repetition rate is set to be 300kHz. Fig.$\ref{fig3}$ shows the image when the laser with different power is shot on the surface. The focusing spot is chosen to be in the middle of periphery and the center\cite{bachman2013permanent,bachman2016threshold}.  Only a small focusing spot can be seen when the laser power is under threshold as shown in Fig.$\ref{fig3}$(a). After 10000 pulses we shut down the femtosecond laser and measure the transmission spectra. Then we raise up the laser power and do the measurement again. The threshold for LN film is measured to be 1.28mW as shown in Fig.$\ref{fig3}$(b). We can clearly see the light is scattered into the cavity and be confined in the periphery. When we shut down the laser a hardly visible defect can be found on the surface. It helps the laser scatter into cavity and the pulses will circulate around the periphery. As the laser power goes up more light will be coupled into the disk and the scattered light will be brighter as can be seen in Fig.$\ref{fig3}$(c) and (d). During the whole process the tapered fiber is attached to the micro-disk in order to keep a fixed coupling condition so that comparisons between different laser powers can be made.

\begin{figure}[b]
    \centering 
    \includegraphics[width=0.95\columnwidth]{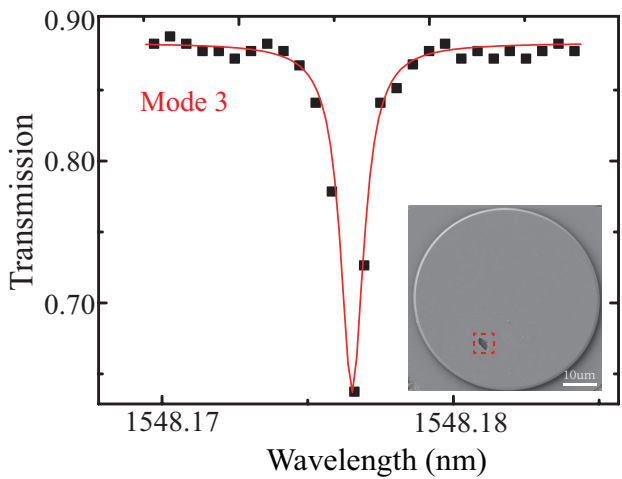}
    \caption{Lorentzian fit of mode 3 centered around 1548.12nm after 16mW femtosecond laser irradiation. The inset shows the defect at half radius place.}
    \label{fig4}
\end{figure}

 The Q factor of some specific mode will experience an increase. Fig.$\ref{fig4}$ (a) shows the Lorentzian fit of mode 3 after 16mW laser irradiation. The intrinsic Q is calculated to be 1.84$\times$10$^6$, which has one order of magnitude increment compared with the original one. The inset shows the defect created by femtosecond laser on LN disk surface.

 \begin{figure}
    \centering 
    \includegraphics[width=0.9\columnwidth]{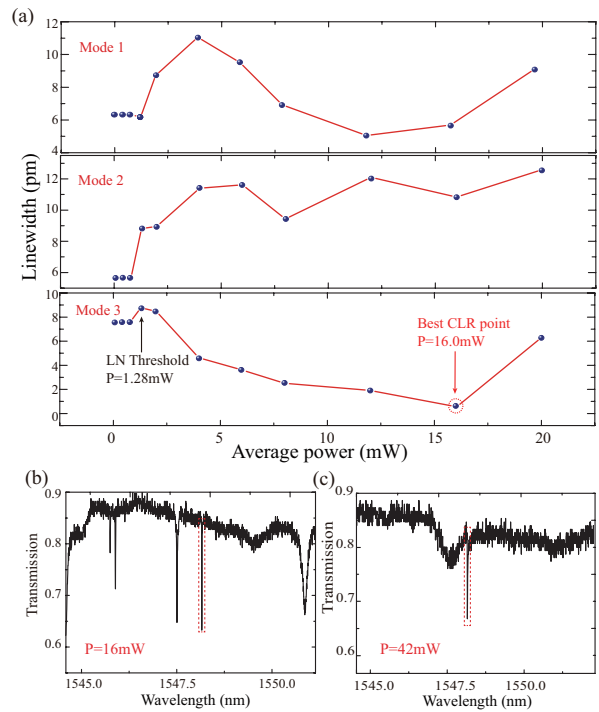}
    \caption{(a) The relationship between linewidth of three marked modes and irradiating laser power. The threshold power for LN film is measured to be 1.28mW while the power for best CLR process is 16mW. (b) Transmission spectra after 16mW laser irradiation. (c) Transmission spectra after 42mW laser irradiation. Mode 3 is marked in a red box.}
    \label{fig5}
\end{figure}

 We also inspected mode 1 and 2 under the same condition, the Q factors of them both have dropped a little to 2.41$\times$10$^5$ and 1.80$\times$10$^5$ respectively. Fig.$\ref{fig5}$(a) shows the relationships between the measured linewidth of the three modes and the irradiating laser power. When the laser power is under threshold, it makes no change on the micro-disk so that the linewidth of all the three modes keeps the same. As soon as the laser power reaches LN threshold, it forms a defect on the surface. The linewidth of mode 1 and 2 experiences an oscillation while modes 3 experiences a continuous decrease as laser power goes up to 16mW. During this process the coupled laser has some interaction with the LN lattices on the periphery which is like a precision surgery that affects optical modes in three aspects. Firstly the heat affected zone induces by femtosecond laser may block optical paths of some high order modes as it extends to the edge of the micro-disk. In this case the Q factors of these modes may have a sudden decrease. Secondly the debris created in the defect formation process may lower the surface cleanliness and thus contributing to a decrease of Q. This could be the main reason for a small drop of Q factor at the threshold. Another effect we suppose is that the defect will help the laser scatter into the cavity\cite{liu2018end,jiang2017chaos} and induce CLR process. Since the laser has high repetition rates and peak power so when it circulates around the cavity it will have some physical effects on the LN lattice\cite{nejadmalayeri2007rapid}. The nonlinear absorption of the laser radiation will result in the formation of ultra-fast electron-hole plasma confined in the solid. Energy will be transferred to lattice via electron-photon coupling, provoking a temperature increase and local melting\cite{gattass2008femtosecond}. Some lattice disorders and tiny burrs on the periphery will be repaired by the laser induced resolidification process after irradiation\cite{garcia2016melt,seuthe2017structural,bonse2004dynamics}. This CLR process is permanent and contributes to an increase of Q factor. We think it is the main reason for Q factor increment of mode 3. When we further increase laser power the defect will be larger and the debris will accumulate on the surface. Both of this will induce a decrease as can be seen in Fig.$\ref{fig5}$(a) when the laser power reaches 20mW. So the linewidth of three modes behaves differently due to the competition between different mechanisms. Fig.$\ref{fig5}$(b) and (c) show the evolution of FSR when we raise up laser power. Fig.$\ref{fig5}$(b) is the FSR after 16mW laser irradiation while Fig.$\ref{fig5}$(c) reveals the FSR after 42mW laser irradiation. Mode 3 is marked in a red box. We can see that all the other modes are disappeared and only mode 3 survived exhibiting relatively high Q factor  when laser power is large enough. We think mode 3 is more localized in the periphery and more sensitive to the changes of the periphery lattices. At the same time femtosecond laser HAZ and surface cleanness have the least influence on it.

In conclusion we have explored a new method to improve Q factor of specific modes in a micro-disk by femtosecond laser irradiation. The pulses confined in the edge of the disk will have compact nonlinear interaction with the lattices which may induce melting or resolidification process. Both of them will help rebuild the periphery and thus improve the quality. We have observed the intrinsic Q factor for a specific mode to have an order of magnitude increment. This CLR process is highly controllable which is very much like a precision surgery and thus offers more possibilities of mode engineering.  In terms of application, the method provides an access to work on a single component without touching other building blocks of an already fabricated chip. It is an easy way to further increase device functionality and may find more roles in integrated chips. This method can improve the quality factor of fundamental modes and suppress other high order modes at the same time, which may provide a more effective and controllable platform for nonlinear optical interactions.

The device was fabricated in Center for Advanced Electronic Materials and Devices (AEMD). This work is supported by National Key R $\&$ D Program of China (2017YFA0303700) and the National Natural Science Foundation of China (NSFC) (11574208).


%
%

%


\bibliography{ref}

\end{document}